\documentclass[useAMS,usenatbib]{mn2e}

\usepackage{placeins}
\usepackage{graphicx}
\usepackage{amsmath}
\usepackage{rotating}
\usepackage{cite}

\newcommand\be{\begin{equation}}
\newcommand\en{\end{equation}}

\newcommand\etal{{\rm et al}.\ }

\title[Dust Feedback to Vortices]{Survival and Structure of Dusty Vortices in Protoplanetary Discs}
\author[I. Crnkovic-Rubsamen, Z. Zhu, and J. Stone]{Ivo Crnkovic-Rubsamen$^{1}$\thanks{E-mail:
ivoc@princeton.edu(IC); zhzhu@astro.princeton.edu(ZZ)}, Zhaohuan Zhu$^{1,2}$, and James M. Stone$^{1}$\\
$^{1}$Department of Astrophysical Sciences, 4 Ivy Lane, Peyton Hall, Princeton University, Princeton, NJ 08544\\
$^{2}$Hubble Fellow}
\begin{document}

\date{}

\pagerange{\pageref{firstpage}--\pageref{lastpage}} \pubyear{2014}

\maketitle

\label{firstpage}

\begin{abstract}
We have studied the impact of dust feedback on the survival and structure of vortices in protoplanetary discs using 2-D shearing box simulations with Lagrangian dust particles. We consider dust with a variety of sizes (stopping time $t_{s} =$ 10$^{-2}\Omega^{-1}$ -- 10$^{2}\Omega^{-1}$), from fully coupled with the gas to the decoupling limit.
 We find that a vortex is destroyed by dust feedback
when the total dust-to-gas mass ratio within the vortex is larger than 30-50\%, independent of the dust size. The dust distribution can still be asymmetric in some cases after
the vortex has been destroyed. 
With smaller amounts of dust, a vortex
can survive for at least 100 orbits, and the maximum dust surface density within the vortex can be more than 100 times larger than the gas surface density, potentially facilitating planetesimal formation. On the other hand, in these stable vortices, small ($t_{s}<\Omega^{-1}$) and large ($t_{s}\ga \Omega^{-1}$) dust grains concentrate differently and affect the gas dynamics in different ways. The distribution of large dust is more elongated than that of small dust. 
Large dust ($t_{s}\ga \Omega^{-1}$) concentrates in the centre of the vortex and feedback leads to turn-over in vorticity towards the centre, forming a quiescent region within an anticyclonic vortex. Such a turn-over is absent if the vortex is loaded with small grains. 
We demonstrate that, in protoplanetary discs where both large and small dust grains are present and under the right condition, the concentration of large dust towards the vortex centre can lead to a quiescent centre,
repelling the small dust and forming a small dust ring around the vortex centre. Such anticorrelations between small and large dust within
vortices may explain the discrepancy between ALMA and near-IR scattered light observations in the asymmetric region of transitional discs. 

\end{abstract}

\begin{keywords}
circumstellar matter -- infrared: stars.
\end{keywords}

\section{Introduction}
One obstacle for planet formation in the core accretion scenario
is the growth of solids from millimeter sized grains to kilometer planetesimals within the  lifetime of protoplanetary discs. Various ideas have been
proposed, including dust collisional coagulation (Weidenschilling 1980, 1995; Blum \& Wurm 2008 for a review),
the gravitational instability for dust layers (Safronov \& Svjagina 1969; Goldreich \& Ward 1973),  the streaming instability (Youdin \& Goodman 2005;
Youdin \& Johansen 2007, Johansen \& Youdin 2007),
and particle trapping in large scale disc structures, such as vortices 
(Barge \& Sommeria 1995; Lyra \etal 2009; Johansen \etal 2004; Heng \& Kenyon 2010; Meheut, Keppens \& Kasse 2012).

Vortices are often the outcome of the hydrodynamic instabilities in rotating flows. 
In protoplanetary discs, many instabilities have been proposed to generate vortices, such as 
the Papaloizou-Pringle Instability (Papaloizou \& Pringle 1984, 1985;
Goldreich \etal 1986, Goodman \etal 1987), the Rossby wave 
instability (RWI) (Lovelace \etal 1999; Li \etal 2000, 2001; de Val-Borro et al. 2006; Lin \& Papaloizou 2010; Lin 2012a, 2012b, Lovelace \&
Romanova 2013 for a review), 
 the baroclinic instability (Klahr \& Bodenheimer 2003;
Petersen \etal 2007a, 2007b; Lesur \& Papaloizou 2010), Vertical Shear Instability (Urpin 
\& Brandenburg 1998; Nelson \etal 2013)
instability at buoyancy critical layers (Marcus \etal 2013),
 convection (Godon \& Livio 2000),
and decaying turbulence (Bracco \etal 1999; Shen \etal 2006). 

Two-dimensional simulations on protoplanetary discs have shown that
anti-cyclonic vortices are long-lived (Godon \& Livio 1999),
and that such vortices can concentrate dust particles significantly (Barge \& Sommeria 1995; Adams \& Watkins 1995; Tanga \etal 1996;
 Godon \& Livio 2000), perhaps thereby  facilitating planetesimal formation. 
 
However, as dust concentrates in the vortex, it becomes more and more critical to consider the dust feedback. When the dust-to-gas mass ratio approaches one, it can lead to vortex destruction (Johansen, Andersen \& Brandenburg 2004). Earlier simulations by Johansen \etal (2004) only evolved the vortex for one orbit and the vortex did not reach a steady state. In order to concentrate dust significantly, the vortex needs to be long-lived. However, when dust is gradually concentrated towards the centre of a vortex the feedback becomes stronger and the vortex can potentially be destroyed. During the preparation of this manuscript, Fu \etal (2014) has shown that dust feedback can significantly shorten the lifetime of the vortex up to a factor of 10.  Raettig et al. (2015) have also investigated the stability conditions, but that work used convective overstability (Klahr \& Hubbard 2014) to generate vortices.

In this paper, we carry out a systematic study of dust feedback on vortex structure and survival using hydrodynamical simulations for up to 100 orbits. 
The feedback from various sized dust with different dust-to-gas ratio is studied. In \S 2, we introduce our method.
The results are presented in \S 3. After a short discussion in \S 4, our conclusions are given in \S 5. 

\section[]{Method}
The gas disc was simulated with Athena (Stone \etal 2008), a higher-order
Godunov scheme for hydrodynamics and magnetohydrodynamics using the piecewise 
parabolic method (PPM) for
spatial reconstruction (Colella \& Woodward 1984), and the corner transport upwind (CTU) method 
for multidimensional integration (Klein, Colella \& McKee 1990)
(Gardiner \& Stone 2005, 2008). Dust particles were simulated with the particle integrator in Athena (Bai \& Stone 2010).
Dust particles were implemented as Lagrangian
particles with the
dust-gas coupling drag term, following
\begin{equation}
\frac{d\mathbf{v}_{i}}{dt}=\mathbf{f}_{i}-\frac{\mathbf{v}_{i}-\mathbf{v}_{g}}{t_{s}}\,,\label{eq:motion}
\end{equation}
where  $\mathbf{v}_{i}$ and $\mathbf{v}_{g}$ denote the velocity vectors for particle $i$
and the gas, $\mathbf{f}_{i}$ is the gravitational force experienced by
particle $i$, and $t_{s}$ is the stopping time for this particle due to gas drag. The dimensionless
stopping time $T_{s}$ is defined as $t_{s}\Omega$.
In this work, we chose the triangular-shaped cloud
(TSC) interpolation scheme to interpolate the gas properties and dust feedback force at the particle position. 
We use the semi-implicit scheme to integrate particle orbits. 
\\
\indent The simulations were conducted using the shearing box approximation. 
The box extends from -0.5 H to 0.5 H with 256 cells in the x-direction (r-direction) and -H to H with 512 cells in the y-direction ($\theta$-direction), where H is the disc scale height ($c_{s}/\Omega$). 
We tested this choice of resolution by running diagnostic tests at double the resolution (512 by 1024) and observed no significant change, leading us to believe that our current resolution was sufficient. 
We applied periodic boundary conditions in the y-direction. To avoid wave reflection from x boundaries, densities and velocities of ghost zones at x boundaries
were fixed at the initial values (as in Zhu \etal 2014).
We set up the Kida vortex by initializing the disc structure as
\begin{align}
v_{x}&=\Omega_{v}y/\chi\nonumber\\
v_{y}&=-\Omega_{v}x\chi\,,\label{eq:vortex2}
\end{align}
for the region within the ellipse described by $\Delta x$
and $\chi\Delta x$ as the semi-minor and semi-major axes. In the
Kida solution, $\Omega_{v}=3\Omega_{0}/(2(\chi-1))$ (Chavanis 2000).
Beyond this region, we assumed that the difference between the speed given by 
(\ref{eq:vortex2}) and the Keplerian speed decreases exponentially
as exp(-$(f-1)$) where $f=(x^{2}/\Delta x^{2}+y^{2}/(\Delta x\chi)^{2})^{1/2}$ is the distance between the vortex centre and the vortex streamline in the $x$ direction. In this study we choose $\Delta x=0.06$ H and $\chi$=4.  Lyra \& Lin (2013) has a good discussion of the difference between the Kida solution and the Goodman--–Narayan--–Goldreich solution.  We note that for our choice of $\chi$, the solutions are quite similar.  We initially applied a small viscosity ($\nu=10^{-5}$) to stabilize
the vortex by damping any oscillations or sound waves generated by discretizing the incompressible Kida solution for use in our compressible grid. After 10 orbits when the vortex is stable, the viscosity was set to be 0. All runs were conducted to 100 orbits. 

The main set of simulations were conducted to explore the parameter phase space given by varying the ratio of solid mass to gas mass, \(\varepsilon\), in the disc and the stopping time of the particles. 
We chose seven $\varepsilon$ from 0.01 to 1 ($\varepsilon=$0.01, 0.0215, 0.0464, 0.1, 0.215, 0.464, and 1), representative of regions of the protoplanetary disc that had already been subject to some solid concentration, either by vertical or radial transport. The
lower limit \(\varepsilon = 10^{-2}\) is the ISM dust-to-gas mass ratio. The upper limit \(\varepsilon = 1\) represents a vortex in a region that has already undergone significant solid concentration or even streaming instability.
\\
\indent We also explored the parameter phase space along the stopping time dimension. Barge and Sommeria (1995) showed that the stopping time is the single parameter that characterizes particle
concentration in the vortex. The stopping time is dependent on the ratio of mass and the size of the particle. Our exploration of stopping time was centered on  $T_{s}$ = 1. We extended this dimension of the phase space \(10^{4/3}\) over 3 simulations in each direction from 0.0316 to 31.6, which captured the limiting behavior in either direction. \\
\indent Thus, we have conducted 7$\times$7 (49) simulations to explore our phase space, varying the mass ratio and stopping time respectively. 

\section{Results}
\subsection{Initial Stability Assessment}
\begin{figure*}
\centering
\includegraphics[width=1.2\textwidth, angle =270]{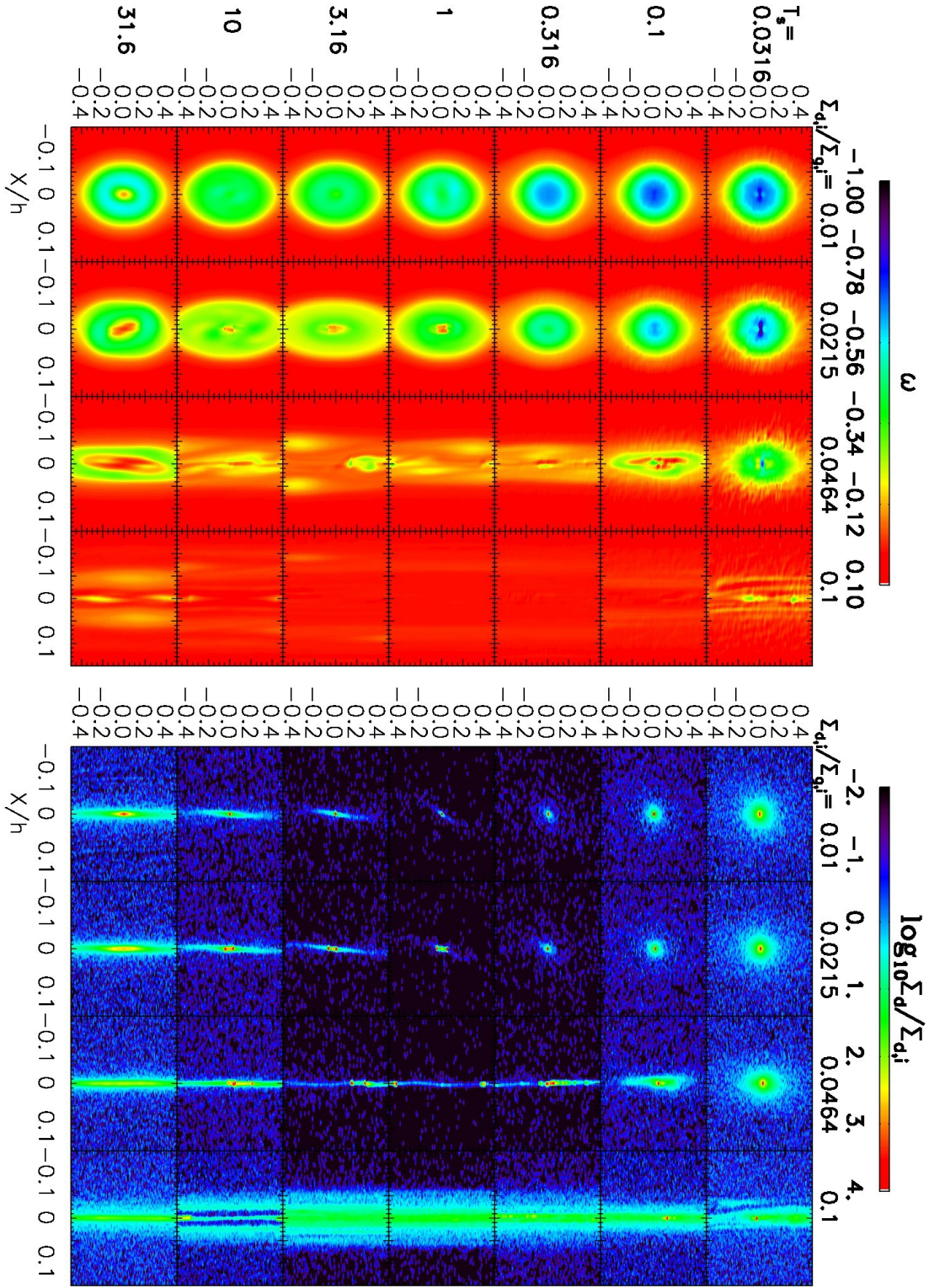} 
\vspace{-0.4 cm}
\caption{Left panel: the vorticity of discs with different initial dust-to-gas mass ratios (columns) and different dust stopping times (rows) at 100 orbits. The feedback from small dust leads to a vorticity minimum at the vortex centre while the feedback from large dust leads to a vorticity turnover at the vortex centre. Right panel: the ratio between dust surface density at 100 orbits and initial dust surface density
for different cases. The distribution of large dust is more elongated than that of the small dust.
} \label{fig:fig1}
\end{figure*}

The long-term stability of vortices for each set of parameters is shown in Fig. \ref{fig:fig1}. The left panel shows the vorticity of all runs at 100 orbits. Since all vortices with $\varepsilon>0.1$ are destroyed after 100 orbits, we do not show those cases in this plot.
Throughout the paper, the vorticity
is defined as $\omega_{z}=(\nabla\times \mathbf{v})_{z}-1.5 \Omega_{0}$, so that the vorticity
due to the Keplerian shear has been removed. When the dust to gas mass ratio increases, there is a change in the stability of vortices near $T_{s} = 1$. At high $\varepsilon$, only vortices with very large or small $T_s$ are stable. At very low stopping times the dust and gas are totally coupled, while for very large stopping times the dust and gas are uncoupled. Both extremes engender stability. However, particles with $T_{s} \approx 1$ sink to the vortex
centre fastest (Chavanis 2000) and these vortices are the least stable, requiring $\varepsilon \leq  0.0215$ for stability. This indicates that for vortices in accretion discs with higher solid mass density, stability is heavily constrained.

Even in cases in which the vortex is destroyed by the dust feedback, there is a region of high dust concentration. This high dust concentration
is produced by the vortex before it is destroyed by feedback. Thus, even if the vortex is a transient phenomenon, it can still lead to very high dust concentration.
	With a small initial dust to gas mass ratio, the vortex can survive over 100 orbits independent on the size of the particle in the disc. However, the structure of the vortex
does depend on the size of the particle within the vortex. With small particles ($T_{s}<1$) in the vortex, the vortex centre has a minimum vorticity. With big particles ($T_{s}>1$)
in the vortex, the vorticity starts to increase towards the vortex centre. This difference has important applications on dust distribution in vortices, which will be discussed in \S 4.

For the smallest particles ($T_{s}=0.0316$) in a stable vortex (the uppermost panel of Fig. \ref{fig:fig1}), we have seen ``fingers'' at the vortex edge, indicating 
some instability for well-coupled particles. We notice that in a recent paper by Fu \etal (2014), a similar phenomenon has been observed. This instability may be due to
the ``heavy core instability'' suggested by Chang \& Oishi (2010) or parametric instability for dust-laden vortices suggested by Railton \& Papaloizou (2014).

\subsection{Stable Vs. Unstable vortices}
\begin{figure*}
\centering
\includegraphics[width=1\textwidth]{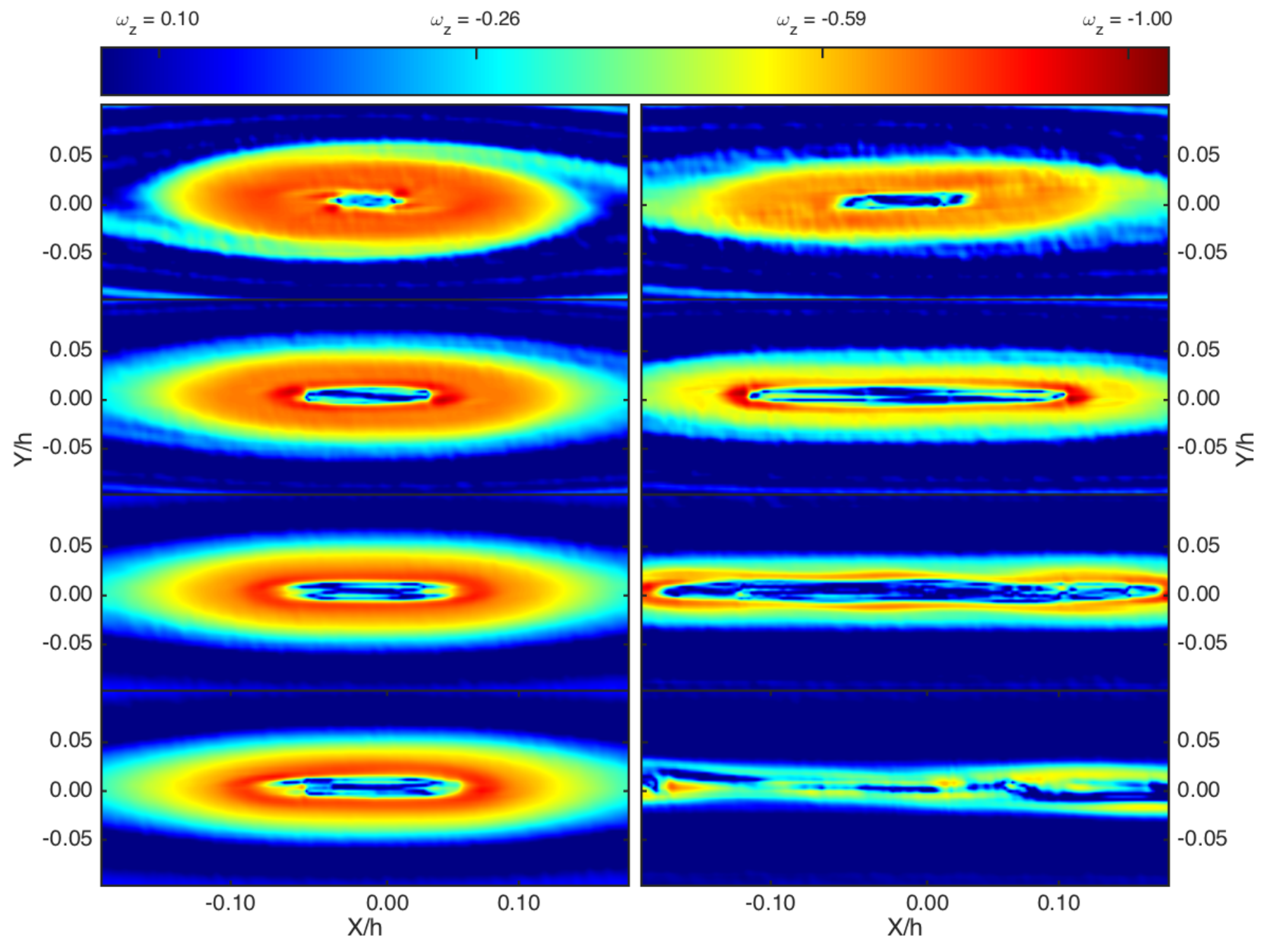} 
\vspace{-0.4 cm}
\caption{Vorticity evolution in a stable (left panel) and unstable (right panel) vortex. The frames are at 5, 10, 20 and 30 orbits, descending from top. The stable vortex is from a simulation with mass ratio $= 0.0215$ and $T_s = 1$, while the unstable vortex has mass ratio $= 0.0464$ and $T_s = 1$. Only the midsection of the shearing box is pictured in each frame for clarity.
} \label{fig:stablevsunstable}
\end{figure*}

We have studied the time evolution of the vorticity of an unstable vortex to understand how decay proceeds. As can be seen in Fig. \ref{fig:stablevsunstable} the vortex quickly starts to stretch in the y direction until it is strongly deformed. This stretch is a consequence of the vortex's own rotational motion becoming slow compared to the shear flow, which is constant through the life of the vortex. Once its own vorticity is no longer large compared to the shear flow, the vortex is simply pulled apart by the shear flow, whereas a stable vortex eventually develops a hole as it fills its core with dust but keeps spinning relative to the shear flow. For dust with $T_{s}>1$, the regions of largest spin are at mid-radial distances, as the vortex slows down around its dust core. Thus, we conclude that even initially stable vortices cannot continue to increase the amount of dust they accumulate indefinitely, or they will approach the critical value of dust to gas ratio identified in \S 3.3 and destroy themselves.

\subsection{Instability Mechanism Diagnosis}
\begin{figure*}
\centering
\includegraphics[width=0.8\textwidth]{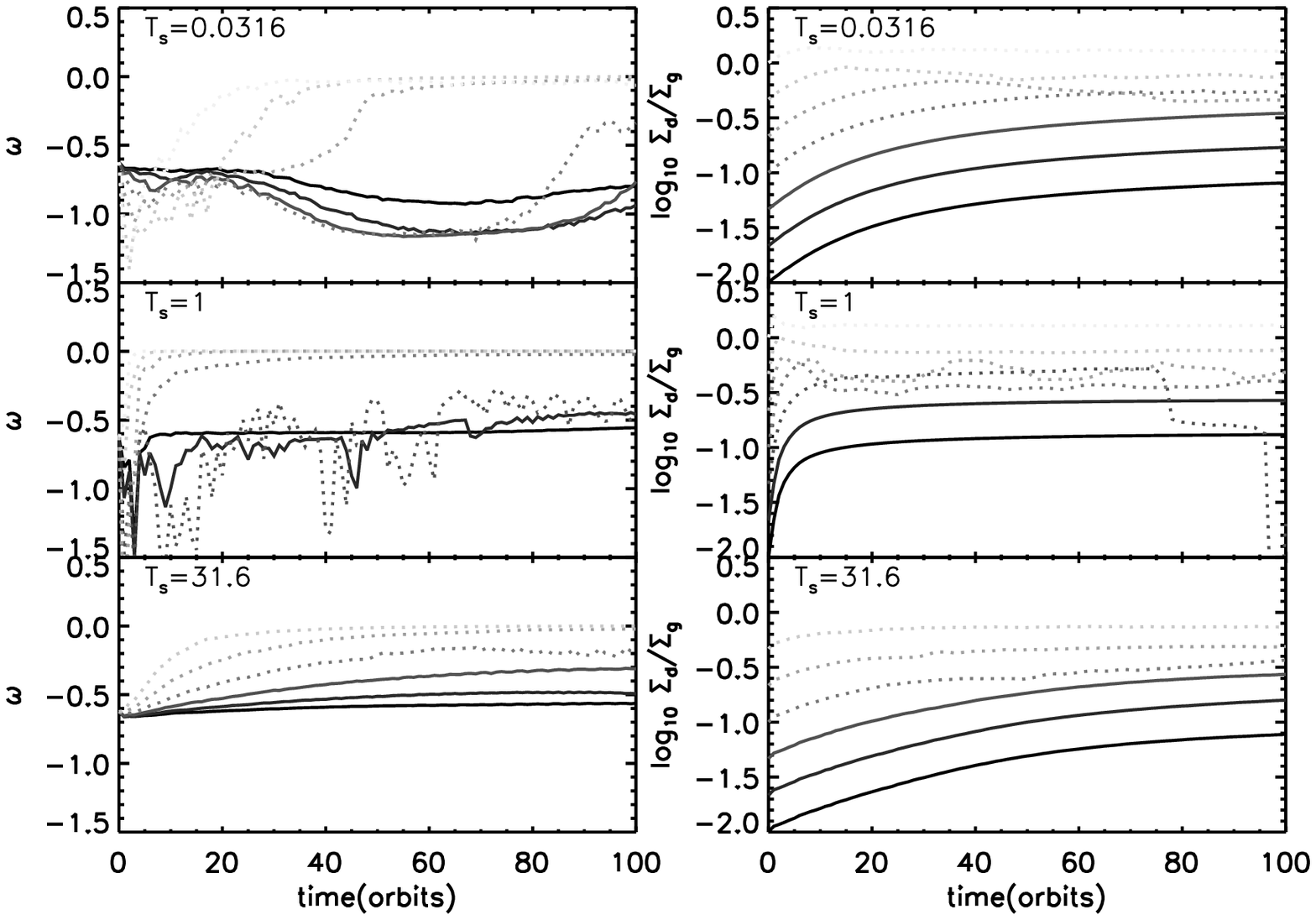} 
\vspace{-0.2 cm}
\caption{Left panels: the minimum vorticity within the vortex for discs having different sized dust particles.
In each panel, darker curves represent the cases with smaller initial dust-to-gas mass ratios. The solid curves
represent the cases where the vortices are still present at 100 orbits while the dotted curves represent the cases
where vortices are destroyed within 100 orbits. Right panels: the averaged dust-to-gas mass ratio within the vortex
region for  discs having different sized dust particles. The curves' color and type have the same meaning as the left panel. 
The vortex is present in cases when the averaged dust-to-gas mass ratio within the vortex is smaller than 0.3. 
} \label{fig:denvort1D}
\end{figure*}

As particles concentrate within the vortex, the dust feedback weakens and eventually destroys the vortex. To quantify the weakening of the
vortex, Fig. \ref{fig:denvort1D} shows the minimum vorticity (left panels) and the averaged dust-to-gas mass ratio (right panels)
within the vortex region. The averaged dust-to-gas mass ratio within the vortex is calculated by summing all the dust mass within the vortex region and dividing this total dust mass
by the total gas mass in the same region. The vortex region is defined to be an ellipse with a semi-major axis of 0.4 H and a semi-minor axis of 0.1 H around the vortex centre. 
In each panel, darker curves represent cases with smaller initial dust-to-gas mass ratios. The solid curves
represent the cases where the vortices are still present at 100 orbits while the dotted curves represent the cases
where vortices are destroyed within 100 orbits. Clearly, when the vortex is destroyed, the vorticity reaches zero quickly. Again,
we have observed that the feedback from small dust (e.g., $T_{s}$=0.0316) leads to a smaller vorticity at the vortex centre while 
the feedback from large dust (e.g., $T_{s}$=31.6) leads to a bigger vorticity. 

Fig. \ref{fig:denvort1D} also suggests that the survival of the vortex seems to depend only on the averaged dust-to-gas mass ratio within the vortex, independent of the dust stopping time. For all three $T_{s}$ presented in Fig. \ref{fig:denvort1D}, the vortex survives over 100 orbits when the averaged 
dust-to-gas mass ratio within the vortex (right panels) is smaller than 10$^{-0.5}\sim$0.3, while larger dust-to-gas mass ratios lead to vortex destruction. 

\section{Discussion}
\subsection{Range of Dust Sizes}
We have identified important differences in the feedback caused by small and large dust particles. In a real protoplanetary disc, small and large dust coexist in the disc. 
However, the gas will be primarily affected by certain sizes of dust. The feedback force from a certain sized dust to the gas is the product of the dust abundance
and the drag force of each dust particle.
Generally, a dust species with a higher abundance will have a bigger effect on the gas. On the other hand, dust with large stopping time will not affect gas significantly. Overall, when the gas structure has been altered by the feedback from the most important dust species, the modified gas structure will affect the distribution of dust at other sizes. 

Assuming ISM dust distribution $dn(s)/ds \approx s^{-3.5}$ in protoplanetary disks, we know that large dust 
normally dominates most of the dust mass. If this large dust has $T_{s}>1$, it can make the vorticity increase towards the vortex centre (Fig. \ref{fig:fig1}).
With a vorticity close to zero, the vortex centre becomes quiescent and small particles cannot concentrate to the vortex centre. Instead, they
will pile up around the vortex centre to form a ring structure. Such anti-correlation between small and big dust may be able to explain the discrepancy of azimuthal
structure between submm and near-IR scattered light observations for Oph IR 48 (Follette \etal 2015).

\begin{figure*}
\centering
\includegraphics[width=1.\textwidth]{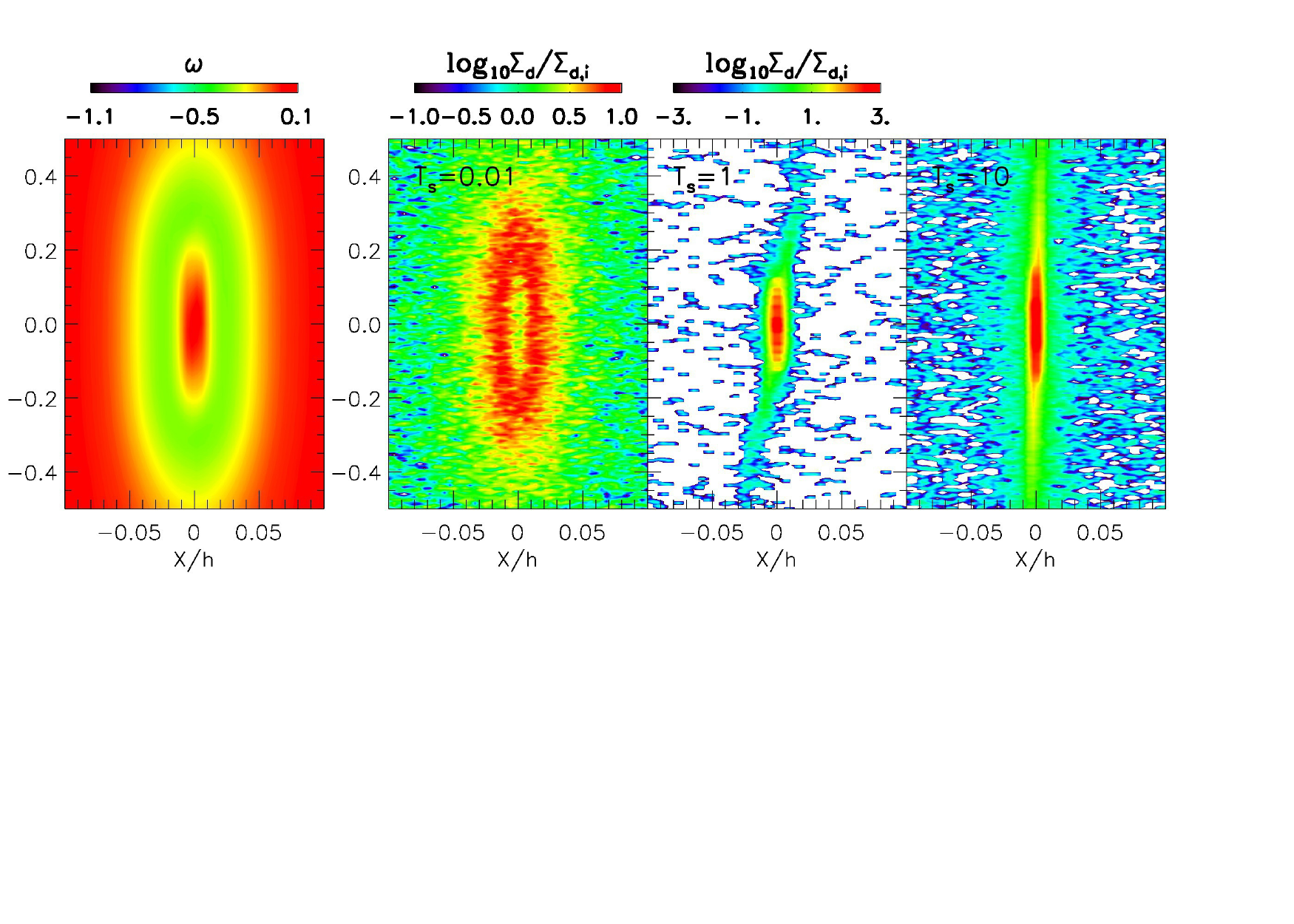} 
\vspace{-5.3 cm}
\caption{The vorticity (the leftmost panel) and  dust distribution (right three panels) from the simulation with particles having a large range of sizes from $T_{s}$=0.01
to 31.6 with a size distribution of $dn(s)/ds\propto s^{-3}$. Since large dust concentrates to the vortex centre, the dust feedback makes
the vorticity increase to 0 at the vortex centre. This makes small particles form a ring
around the vortex centre. 
} \label{fig:multipargen}
\end{figure*}

In order to explore this interesting possibility, we have carried out a simulation with particles having a large range of sizes from $T_{s}$=0.01
to 31.6 with a size distribution of $dn(s)/ds\propto s^{-3}$. The total dust-to-gas mass ratio is 0.03. In this case, most of the dust mass is in big particles. The gas vorticity and dust distribution at
50 orbits are shown in Fig. \ref{fig:multipargen}. As expected, since large dust concentrates in the vortex centre, the dust feedback makes
the vorticity approach 0 at the vortex centre. This prevents small dust from concentrating in the vortex centre. Instead, the small particles form a ring
around the vortex centre. The vortex is gradually weakened by the feedback, and as it becomes more elongated, the ring structure of the small particles also elongates. The implications of these particle distributions are discussed in \S 4.2.

\subsection{Particle Distribution Within the Vortex}
\begin{figure}
\centering
\includegraphics[width=0.5\textwidth]{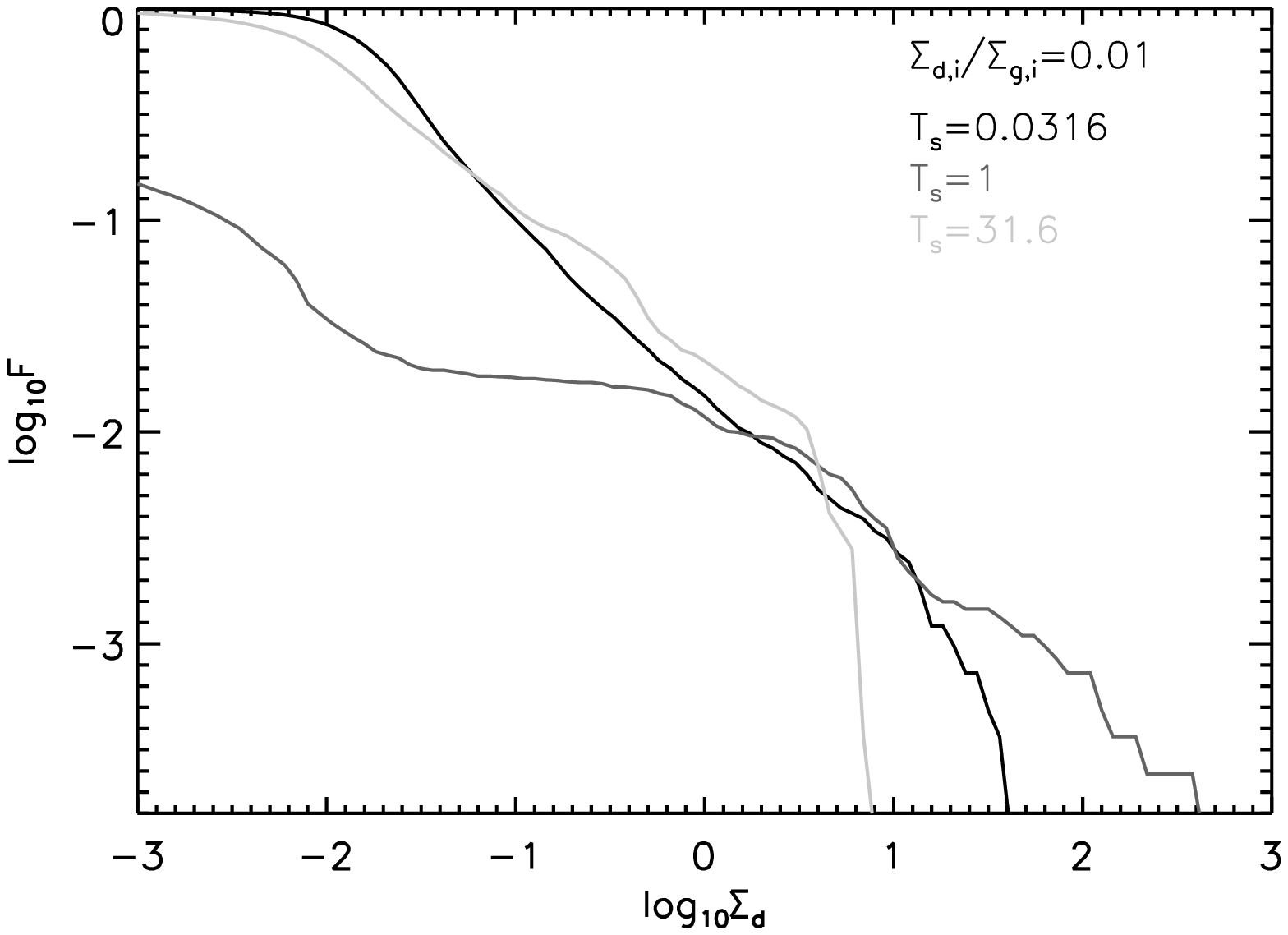} 
\vspace{-0.4 cm}
\caption{The fraction of the vortex region having dust surface density larger than $\Sigma_{d}$ for stable vortices with $T_{s}$=0.0316, 1, and 31.6. 
} \label{fig:pdf}
\end{figure}

By comparing ALMA synthetic images based on simulations with ALMA observations towards Oph IRS 48 (van der Marel et al. 2014) and HD 142527 (Casassus et al. 2013), 
Zhu \& Stone (2014) point out that the vortex centre in these observations may be optically thick. Since in regions that are optically thick, the intensity is no longer proportional to the total dust mass, it is important to know the fraction of the vortex region that is optically thick. 
Thus, we show the dust surface density in Fig. \ref{fig:pdf}. The figure indicates
that the maximum dust density within a stable vortex can reach $>$100, even though this only occurs in a small region around the vortex centre. As expected, particles with $T_{s}\sim$1 reach the largest concentration at the vortex centre. By assigning an opacity to the particles, we know the $\Sigma_{d}$ beyond which the vortex becomes optically thick. Then with Fig. \ref{fig:pdf},
we can estimate how much is hidden in the optically thick region. We also note that such high concentration at the vortex centre can lead to gravitational instability to form planetesimals or even planets (Barge and Sommeria 1995).

\subsection{Vortices with different sizes}
\begin{figure*}
\centering
\includegraphics[width=0.8\textwidth]{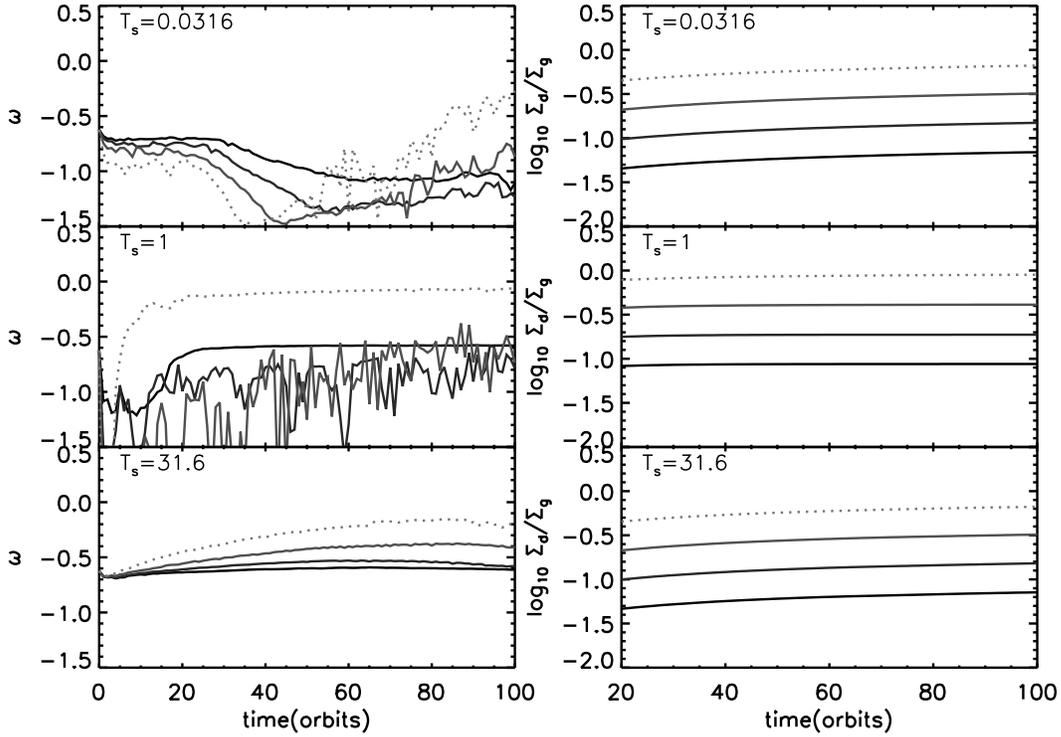} 
\vspace{-0.2 cm}
\caption{Similar to Fig. \ref{fig:denvort1D}, but the vortex is 5 times larger.
Similar to small vortices, the vortex is present in cases when the averaged dust-to-gas mass ratio within the vortex  is smaller than 0.3. 
} \label{fig:denvort1Dbig}
\end{figure*}

In our main set of simulations, we initialized the vortex with a semi-minor axis of 0.06 H. In order to explore
how the vortex size affects our results, we carried out a similar set of simulations with a
vortex 5 times bigger ($\Delta$x=0.3 H). The whole simulation domain was thus adjusted to [-2.5H, 2.5H]$\times$[-5H, 5H]. 
To save the computational cost, we only set the mass ratio ($\varepsilon$) to be $0.01$, $0.0215$, $0.0464$ and $0.1$, and
the dust stopping time ($T_{s}$) to be $0.0316$, $1$, $31.6$,  All other parameters, including grid numbers and simulation time, are the same as our main set of simulations. We again found that vortices with $\varepsilon\ge0.1$ are unstable. Similar to Fig. \ref{fig:denvort1D},
Fig. \ref{fig:denvort1Dbig}
shows that the vortex is destroyed when the averaged dust-to-gas mass ratio within the vortex is larger than $\sim$0.3.

\section{Conclusions}
We have studied dust feedback in vortices in protoplanetary discs using 2-D shearing box simulations with Lagrangian dust particles. 
Dust with a variety of sizes (stopping time $t_{s} =$ 10$^{-2}\Omega^{-1}$ -- 10$^{2}\Omega^{-1}$), from fully coupled with the gas to the decoupling limit, have been considered. 
\begin{itemize}
\item We find that the vortex is destroyed by dust feedback
when the dust-to-gas mass ratio within the vortex is larger than 30-50\%, independent of the dust size.

\item  Even in cases where the vortex is destroyed by dust feedback, we find that there is a region of high dust concentration. This concentration is produced by the vortex before it is destroyed by feedback. Thus, even if the vortex is a transient phenomenon, it can still lead to very high dust concentrations. 

\item We find that small ($t_{s}<\Omega^{-1}$) and large ($t_{s}\ga \Omega^{-1}$) dust concentrate differently in the vortex
and affect the gas vortex structure differently. The distribution of large dust is more elongated than that of small dust. Large dust ($t_{s}\ga \Omega^{-1}$) concentrates in the vortex centre and dust feedback leads to an turn-over in vorticity towards the vortex centre, forming a quiescent centre within the anticyclonic vortex. Such a turn-over is absent if the vortex is loaded with small dust. 
 
\item We have demonstrated that in protoplanetary discs where both small and large dust is present, with the large dust comprising most of the dust mass, the concentration of large dust towards the vortex centre will lead to a quiescent centre, repelling the small dust and forming a small dust ring around the vortex centre. Such anticorrelation between small and large dust within vortices may explain the discrepancy between ALMA and near-IR scattered light observations in the asymmetric region of transitional discs. 

\item We find that the dust density can be more than 100 times larger than the gas density at the vortex centre, potentially facilitating planetesimal and planet formation. 
\end{itemize}

However, we caution that our 2-D simulations have not considered the 3-D structure of the vortex and how vertical transport affects particle distributions.
These questions of 3-D structure deserve future study.  Additionally, we have ignored the radial drift of the particles in our simulations.  We note that the vortices we study tend to appear at density bumps, due to their formation by the Rossby Wave Instability for example.  These density bumps have much reduced or negligible radial drift, so this may limit the magnitude of the effect we're failing to capture. On the other hand, dust radial drift may trigger streaming instability which deserves future investigation. Finally, the formation of real vortices is driven by the presence of a physical instability, which may have significant effects on vortex stability and character.  Conducting simulations which better represent this reality than simply introducing vortices as an initial condition could be a fruitful path for future work.  Refer to Raettig et al. (2015) for a study in this direction.

\section*{Acknowledgments}



\end{document}